# The VELO Upgrade

**Eddy Jans,**[a] (on behalf of the LHCb VELO Upgrade group)

[a] *Nikhef,*
   *Science Park 105, 1098 XG Amsterdam, The Netherlands*
   *E-mail*: `e.jans@nikhef.nl`

ABSTRACT: A significant upgrade of the LHCb detector is scheduled to be installed in 2018-2019. Afterwards all sub-detectors will be read out at the LHC bunch crossing frequency of 40 MHz and the trigger will be fully implemented in software. The silicon strip vertex detector will be replaced by a hybrid pixel detector. In these proceedings the following items are discussed: frontend ASIC, data rates, data transmission, cooling, radiation hard sensors, module design and simulated performance.

# Contents

## 1. Introduction

The LHCb detector [1] is a forward single-arm spectrometer at one of the four collision points of the Large Hadron Collider (LHC) at CERN. The physics programme of the LHCb collaboration is very broad, it ranges from high precision measurements of the production and decays of heavy flavoured hadrons, study of *CP* violation processes, rare decays to production of long lived heavy particles. The detector consists of a vertex detector (VELO), dipole magnet, tracking stations, calorimeters and muon stations. Together they feature good momentum resolution, excellent particle identification and a flexible trigger. During Run 1 of the LHC, which lasted from 2010 till 2013, 3 fb$^{-1}$ of integrated luminosity was collected at an instantaneous luminosity of $\leq 4\times 10^{32}$ cm$^{-2}$s$^{-1}$, which corresponds to an average rate of 1.5 visible interactions per bunch crossing.

During Long Shutdown2 (LS2) of the LHC, which is scheduled to start in Q3 of 2018, the detector will be upgraded such that all sub-detectors are read out at the bunch crossing frequency of 40 MHz, in combination with a fully software-implemented trigger [2]. During Run 3, from 2020 on, LHCb will operate at a 5 times higher instantaneous luminosity. Thanks to a more efficient trigger, this will lead to an increase of the yield by a factor 10 to 20, depending on the process of interest. The layout of the upgraded LHCb detector is shown in Figure 1.

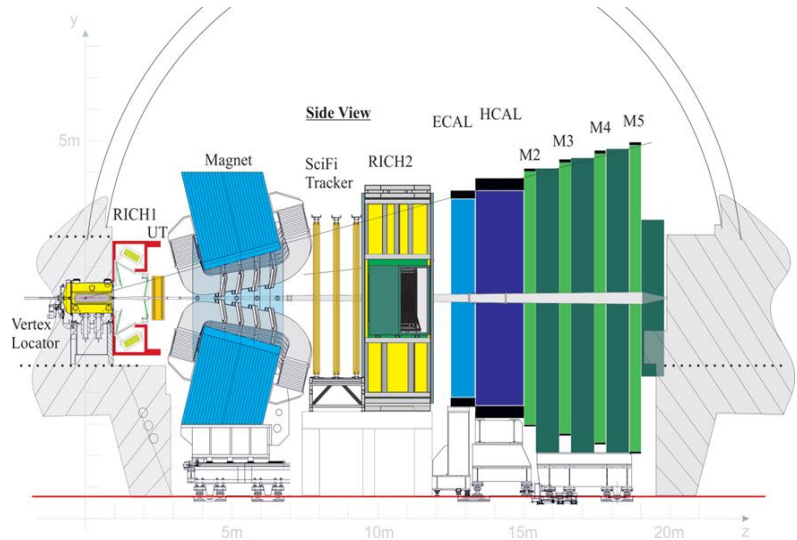

**Figure 1:** Side view of the LHCb detector after the Upgrade.

### 1.1 Present VELO

The main task of the VELO is to enable LHCb to reconstruct and trigger on displaced vertices. The current VELO detector consists of 42 modules each equipped with two silicon strip sensors: with one sensor the radial and with the other the azimuthal angle is measured [3]. The detector is divided in two moveable halves, which allows to retract each of them by 3 cm during injection and tuning of the LHC beams. Once the beam conditions are stable, the halves are closed and centred around the luminous region. The sensors operate in a secondary vacuum, which is separated from the beam vacuum by a 300 μm thick aluminum RF-foil. The first active strip starts at 8.2 mm from the centre of the interaction region. The sensors are cooled by an evaporative cooling system using bi-phase $CO_2$ at -28 $^o$C.

For the Upgrade the present VELO will be replaced by hybrid pixel detectors with planar silicon sensors. Various aspects of the VELO-Upgrade project are discussed in sections 2 and 3.

## 2. VELO Upgrade

Extensive Monte Carlo simulations have resulted in a detector layout with two halves each containing 26 modules. Each module is positioned perpendicular to the beam direction and has an L-shaped active area of ~24 cm$^2$.

### 2.1 Distance to the beams

The impact parameter resolution, $\sigma_{IP}$, is a key performance parameter of the VELO as it quantifies the capability to select events in which beauty and charm hadrons decay. It depends, among other things, on the radius at which the first point of the track is measured and on the amount of material traversed between the primary vertex and the second measured point. Therefore it is advantageous to bring the sensors as close as possible to the interaction region and to minimize the material budget, especially that of the RF-foil.

For the Upgrade the inner radius of the RF-foil will be reduced to 3.5 mm, such that the first pixel will be at 5.1 mm from the interaction region.

## 2.2 Data rates and VeloPix ASIC

The readout of the pixel sensors will be performed by VeloPix ASICs, which are based on the TimePix3 design [4]. An ASIC consists of a matrix of 256x256 pixels, of 55x55 $\mu m^2$ each and therefore covers an area of 198.2 $mm^2$. Each module contains 12 ASICs. The anticipated luminosity of $2 \times 10^{33}$ $cm^{-2}.s^{-1}$ translates for the hottest chip into a rate of $6 \times 10^8$ pixel hits per second. Although the chip has zero-suppression, the corresponding data rates are very high due to the fact that it operates without trigger. In Figure 2 the data rate of each individual ASIC is shown, where the yellow arrows indicate the column wise direction of readout.

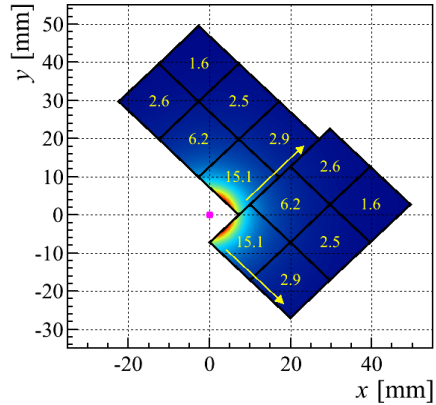

**Figure 2:** Schematic orientation of the 12 ASICs of a module. The red dot at [0,0] represents the colliding beams. The numbers in yellow give the expected data rate per VeloPix chip in Gb/s.

The wafers with the VeloPix chips will be thinned to 200 $\mu m$, before they are diced. More details about the design of the VeloPix chip can be found in [5].

## 2.3 Data transmission

The data of the frontend ASICs is transmitted at 5.1 Gb/s via copper links to feedthroughs on the vacuum tank of the VELO. These kapton cables have to be radiation hard and at the same time flexible enough to permit the VELO halves to move out and in at each fill of the LHC. At the outside of the vessel the electrical to optical conversion takes place. From here 1040 optical links transfer the data stream of 2 TB/s to the counting house where FPGAs on custom made PCIe40 boards perform the decoding and time ordering. Next the events are sent to the trigger farm, where cuts on the data are applied, such that the sample is enriched and reduced to a rate of 20 kHz that is subsequently written to disk. A schematic layout of the data path is shown in Figure 3.

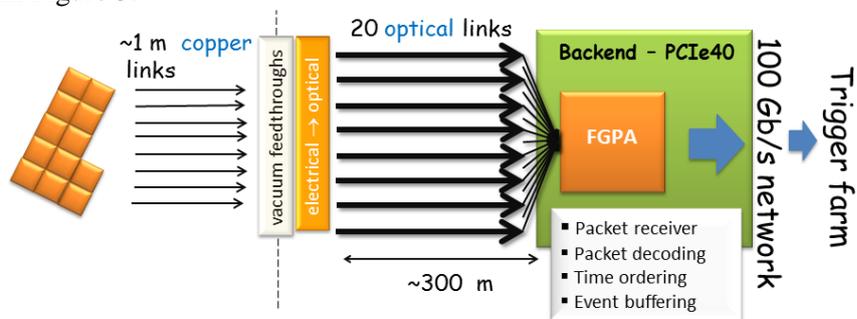

**Figure 3:** Data path from the detector in vacuum to the computer farm on the surface.

## 2.4 Sensors

To minimize dead regions it is decided that three ASICs will be bump-bonded to one sensor and as such form a tile. In the 110 μm wide gap between the chips the sensor has elongated pixels, such that there is no dead area. The layout of a tile is shown in Figure 4.

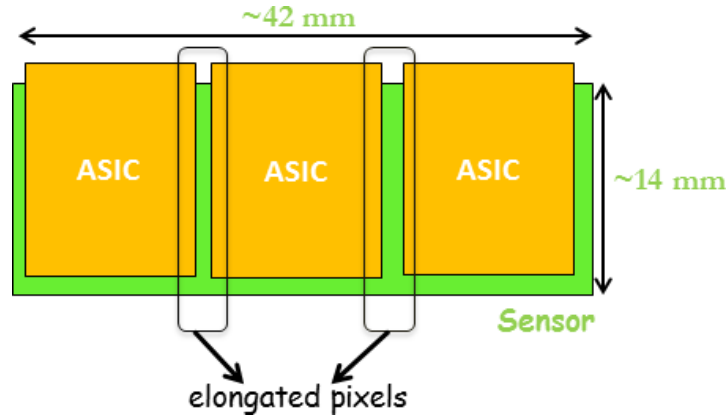

**Figure 4:** Schematic drawing of a tile with a sensor (green) and three VeloPix ASICs (ochre). The ASICs protrude beyond the sensor by ~2 mm to permit wire bonding of their pads.

The luminosity that is planned to be accumulated amounts to 50 fb$^{-1}$, which corresponds to a fluence of 8x10$^{15}$ 1 MeV n$_{eq}$/cm$^2$ at the innermost tip of the sensors. A factor 40 lower fluence is expected at the opposite side of the tile, which is located at ~40 mm from the interaction region. The combination of high dose and inhomogeneous radiation profile pushes the sensor design to the limit. The baseline is 200 μm thick sensors that are capable of sustaining a bias voltage of 1000 V, by using a 450 μm wide multi-guard-ring structure. At present various sensor designs that have been produced by two vendors are under study in the lab and in test beams: n-in-n and n-in-p technology, conservative and more aggressive guard ring designs, etc.

## 2.5 Cooling

Similar to the present VELO the upgraded VELO will use evaporative $CO_2$ cooling to prevent thermal runaway of the sensor assemblies. Hence all parts of the sensor should remain below -20 °C. The expected heat load generated by the sensors (4x) and ASICs (12x VeloPix, 2x GBTx and 2x SCA) is 43 W per module. To fulfill this requirement the novel microchannel cooling technique has been selected [6]. The coolant flows through 16 microchannels of 120x200 μm$^2$, which are etched in a 260 μm thick wafer in such a pattern that it passes directly below the heat producing components. This wafer is subsequently bonded hydrophobically to a cover wafer, such that finally a substrate of 400 μm thickness results. In this way a detector, composed of only silicon parts, is constructed, which is beneficial in view of minimal CTE-mismatch.

In order to reduce multiple scattering before the first measuring point the sensor at the inner edge will protrude beyond the substrate by 5 mm. The additional ΔT that is introduced by this geometry amounts to 5 °C and can be accommodated by supplying bi-phase $CO_2$ at -35 °C. A schematic drawing of the microchannel substrate is shown in Figure 5.

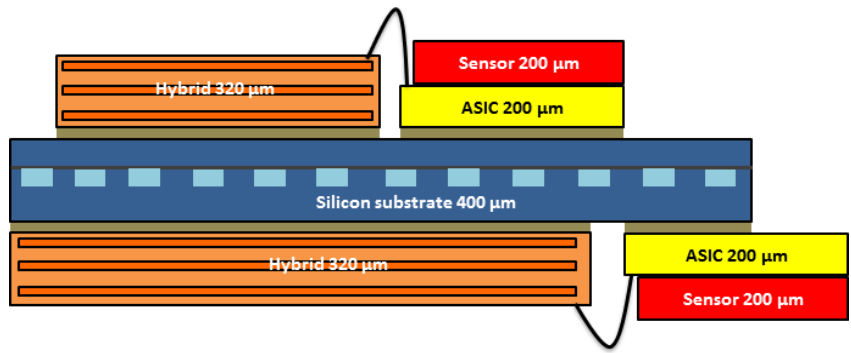

**Figure 5:** Cross section of the silicon microchannel substrate (blue) with on either side tiles (sensor in red and ASIC in yellow) and multi-layer kapton hybrids (brown).

Making a reliable connection between the 0.4 mm thick silicon substrate and the stainless steel capillaries that supply the $CO_2$ is a technological tour de force. More details about this and the microchannel cooling project can be found in [7].

**2.6 Module designs**

Two module designs, both shown in Figure 6, are being developed and tested. They differ in the way the detector part, i.e. the microchannel substrate with on each side a hybrid and two tiles, is connected to the module foot. In the one on the left the substrate is glued on a 1 mm thick carbon fiber plate, which gives good mechanical stability in the transverse directions. However, in the $z$-direction, i.e. along the beam line, the structure is very flexible, such that an additional constraint system will be needed to reach the required positioning precision. The design shown on the right uses 1 mm thick carbon fiber rods with an outer diameter of 6 mm. The 1 mm thick carbon plate between the two rods is glued to the kovar $CO_2$ connector. This forms the only thermal and mechanical connection between the two parts. Prototypes of both designs are being produced and qualified with respect to deformation under mechanical stress and cooling down to -30 $^o$C.

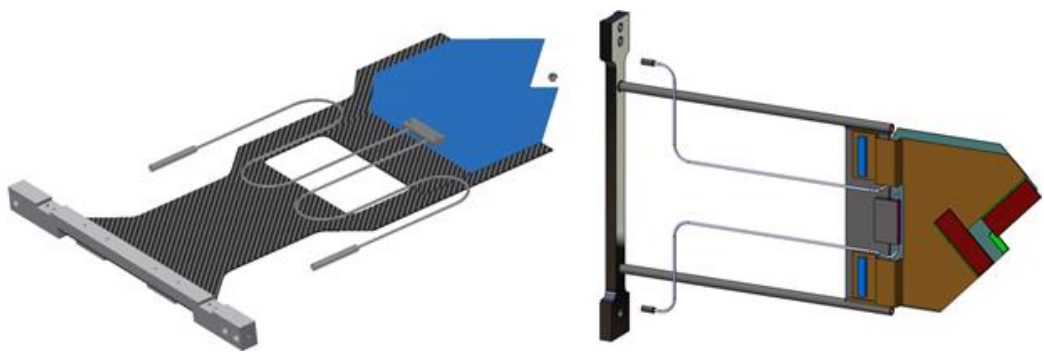

**Figure 6:** The two prototype module designs: (left) with a carbon fiber plate as supporting structure; (right) with two hollow carbon fiber rods.

## 2.7 RF-box

The detectors of a VELO-half are surrounded by an RF-box that separates the ultra-high vacuum of the beam volume from that of the detector. Moreover, they guide the mirror currents and shield the detectors from possible RF pick-up from the beams. The shapes of the sides facing the beams are constrained because the RF-boxes should not contribute significantly to the impedance of the LHC. During the venting and evacuating procedures the foil has to withstand differential pressures of 5 mbar. As the material budget of the VELO is dominated by the contribution of the RF-foil it should be made as thin as possible and preferentially of a material with a high radiation length. The wish to bring the sensors as close as possible to the beams led to the choice of leaving only 0.8 mm clearance between the RF-foil and the nearest point of the sensor. These conflicting requirements put high demands on the choice of material as well as the production process. Extensive tests on prototype boxes have shown that it is possible to meet the demands when the corrugated structure is milled from a solid block of AlMg4.5, provided many precautions are taken into account. A picture of a milled prototype RF-box, which has a quarter of the length of the final version, is shown in Figure 7. Options to further reduce the thickness of the relevant part of the RF-foil, i.e. the strip around the center where the beam passes, are being investigated. The most promising one is to further thin this part of the foil by chemical etching with NaOH.

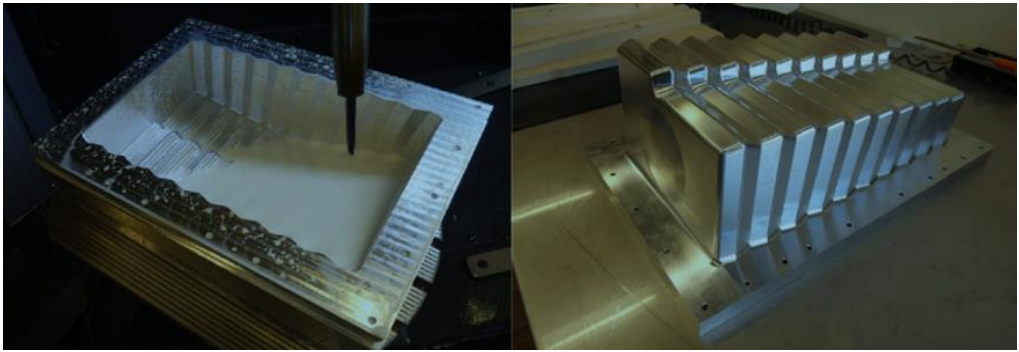

**Figure 7:** Prototype RF-box being milled from a solid block of AlMg4.5 (left) and the 250 μm thick result (right).

## 3. Simulated physics performance

Ray-tracing simulations are performed to optimize the performance by varying the number of modules and their $z$-positions. Finally a layout is chosen in which 26 modules are placed in each detector half at multiples of 12.5 mm along the beam direction [8]. The geometric efficiency for tracks having hits in four planes is 99.95 %. The tracking efficiency as a function of the azimuthal angle ϕ is shown in Figure 8 for the current and upgraded VELO under nominal Upgrade conditions. The more uniform and higher efficiency of the pixel detector is clearly visible.

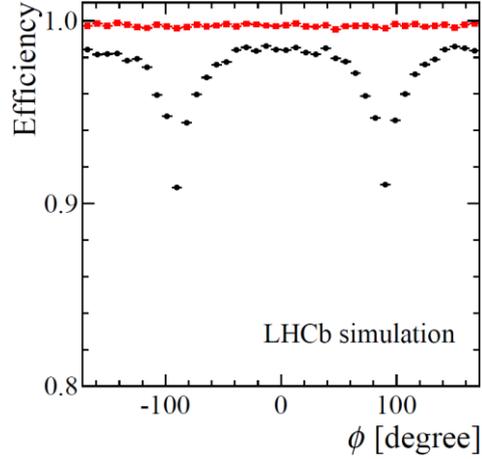

**Figure 8:** Tracking efficiency for current (black) and upgraded VELO (red).

The impact parameter (IP) of a track is the key parameter in the LHCb trigger software to identify long-lived particles like $B$ mesons. In Figure 9 the IP resolution is shown as a function of the inverse transverse momentum ($1/p_T$) for the current and upgraded VELO for nominal Upgrade conditions. In grey the momentum distribution of a typical $B_s$-meson is displayed. For all $p_T$ values the performance of the new pixel detector surpasses that of the present VELO.

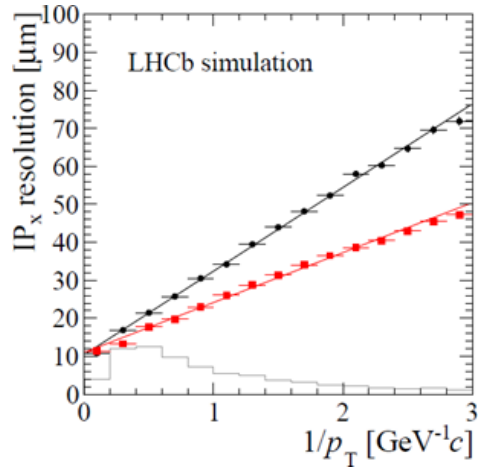

**Figure 9:** Projected impact parameter resolution of the upgraded VELO detector (red) and current VELO (black).

For many of the key measurements of the LHCb physics program, the measurement of $B_s$ meson oscillations is crucial. The corresponding decay time resolution depends on the reaction and is shown in Figure 10 for $B^0 \to K^{*0} \mu^+ \mu^-$. The simulated value is 35 fs which is slightly better than the presently measured one and translates in a smaller statistical uncertainty for the same amount of data.

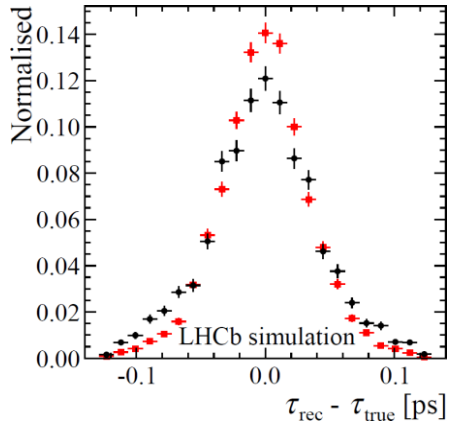

**Figure 10:** Decay time distribution for $B^0 \rightarrow K^{*0}\mu^+\mu^-$ for the upgraded VELO detector (red) and current VELO (black).

## 4. Summary and outlook

The LHCb detector will be upgraded during the second long shutdown of the LHC in 2018-2019. After the Upgrade LHCb will run at a 5 times higher luminosity and the plan is to collect 50 fb$^{-1}$. The VELO will be replaced by a hybrid pixel detector that will be read out at the full bunch crossing frequency. Various technological challenges have already been overcome, while others are under investigation.